\documentstyle[12pt,epsfig]{article}

\def\bseq{\begin{subequation}}  
\def\eseq{\end{subequation}}
\def\bsea{\begin{subeqnarray}}  
\def\esea{\end{subeqnarray}}

\def\bc{\begin{center}}
\def\ec{\end{center}}

\def\bfg{\begin{figure}}
\def\efg{\end{figure}}

\newcommand{\bbox}{\lower.2ex\hbox{$\Box$}}

\newcommand{\beq}{\begin{equation}}
\newcommand{\eeq}{\end{equation}}
\newcommand{\bea}{\begin{eqnarray}}
\newcommand{\eea}{\end{eqnarray}}
\newcommand{\ena}{\end{eqnarray}}

\newcommand {\non}{\nonumber}

\renewcommand{\a}{\alpha}
\renewcommand{\b}{\beta}
\renewcommand{\c}{\gamma}
\renewcommand{\d}{\delta}
\newcommand{\th}{\theta}

\newcommand{\pa}{\partial}

\newcommand{\G}{\Gamma}

\newcommand{\e}{\epsilon}

\renewcommand{\l}{\lambda}

\newcommand{\m}{\mu}

\newcommand{\n}{\nu}

\newcommand{\F}{\Phi}

\newcommand{\p}{\pi}

\newcommand{\s}{\sigma}
\renewcommand{\S}{\Sigma}

\newcommand{\olW}{\overline{W}}
\newcommand{\olcalW}{\overline{\cal W}}
\newcommand{\olD}{\overline{D}}
\newcommand{\olK}{\overline{K}}
\newcommand{\olth}{\overline{\th}}
\newcommand{\olPhi}{\overline{\Phi}}
\newcommand{\oll}{\overline{\l}}
\newcommand{\olf}{\overline{f}}

\newcommand{\calW}{{\cal W}}

\newcommand{\calY}{{\cal Y}}

\newcommand{\ad}{{\dot{\alpha}}}
\newcommand{\bd}{{\dot{\beta}}}
\newcommand{\cd}{{\dot{\gamma}}}

\newcommand{\ulu}{\underline{1}}
\newcommand{\uld}{\underline{2}}

\newcommand{\NP}[1]{Nucl.\ Phys.\ {\bf #1}}
\newcommand{\PL}[1]{Phys.\ Lett.\ {\bf #1}}

\newcommand{\MPL}[1]{Mod.\ Phys.\ Lett.\ {\bf #1}}

\begin{document}

\begin{titlepage}
\begin{flushright} IFUM-644-FT\\ KUL-TF-99/27  \\
\end{flushright}
\vfill
\begin{center}
{\LARGE\bf $(\a')^4$ corrections to the $N=2$ supersymmetric
Born-Infeld action  }\\
\vskip 17.mm  \large
{\bf  A.~De Giovanni $^1$, A.~Santambrogio $^2$ and D.~Zanon $^1$}\\
\vfill
{\small
$^1$ Dipartimento di Fisica dell'Universit\`a di Milano \\
and INFN, Sezione di Milano,
via Celoria 16, I-20133 Milano, Italy\\
\vspace{6pt}
$^2$ Instituut voor Theoretische Fysica,
Katholieke Universiteit Leuven,\\ Celestijnenlaan 200-D,
B-3001 Leuven, Belgium }
\end{center}
\vfill

\begin{center}
{\bf ABSTRACT}
\end{center}
\begin{quote}

We consider the $N=2$ supersymmetric Born-Infeld action and compute
one-loop
divergences quantizing the theory in $N=1$ superspace.
We find that in the presence of non constant curvature the theory is
not
renormalizable. The structure of the $(\a')^4$ counterterm,
proportional to
derivatives of the curvature, is
consistent with effective action calculations from superstring theory.

\vfill
\vskip 5.mm
 \hrule width 5.cm
\vskip 2.mm
{\small
\noindent e-mail: antonio.degiovanni@mi.infn.it\\
\noindent e-mail: alberto.santambrogio@fys.kuleuven.ac.be\\
\noindent e-mail: daniela.zanon@mi.infn.it}
\end{quote}

\begin{flushleft}
July 1999
\end{flushleft}
\end{titlepage}

A nonlinear generalization of Maxwell theory  is given by
the Born-Infeld action \cite{BI}
\bea   \label{SBI}
S_{BI} & = & \frac{1}{\alpha^2} \int d^4x \left\{ 1 -
\sqrt{-det(\delta_{\m\n}+\alpha~F_{\m\n} ) } \right\}
\eea
where we have defined $\alpha\equiv 2\pi \alpha' $ and have chosen a
flat
euclidean metric $g_{\m\n}=\delta_{\m\n}$.
The action in (\ref{SBI}) describes the self-interaction of
electromagnetic fields of arbitrary degree in the curvature tensor
$F_{\m\n}$.
It has the remarkable property of reproducing the
effective action from open bosonic strings in a background abelian
gauge
field with constant curvature \cite{FT,ACNY}.

Given the fact that the Born-Infeld action
contains terms of arbitrary order in the field-strength,
a question that naturally arises is related to its renormalizability
properties: is the theory finite, renormalizable or at least
renormalizable
for constant curvature? In the following we address these issues for
the
supersymmetric version of the theory, computing in superspace at the
one-loop level.

\vspace{0.6cm}

The $N=1$ supersymmetrization of the action in (\ref{SBI}) has been
obtained
in ref. \cite{CF}, while the extension to the $N=2$ supersymmetric case
has
been presented in \cite{ketov}. We focus on the $N=2$ version of the
theory:
it can be obtained by a partial breaking of $N=4$ supersymmetry and
therefore  it is consistent with the effective $D3$-brane action
\cite{APS}
in the type IIB superstring theory.

We introduce the $N=2$
superfield-strength ${\calW}$ satisfying the constraints
\bea
&& \olD_{\ad i} \calW = 0
\qquad \qquad D^4 \calW= \Box\olcalW
\label{constraint}
\eea
with $i=\ulu, \uld$. The $N=2$ superspace invariant action can be
written as \cite{ketov}
\bea
S_{SBI} &=& \frac{1}{2} \int d^4 x d^4 \th ~\calW^2
        + \frac{1}{8} \int d^4 x d^4 \th d^4\olth
        ~\calY (K,\olK) ~\a^2~\calW^2 \olcalW^2
\label{sbi2}
\eea
where $d^4 \th \equiv d^2
\th_{\ulu} d^2 \th_{\uld}$. The function $\calY(K,\olK)$ defined in
terms
of $ K= D^4 \calW^2 $
is given explicitly by
\bea
\calY(K,\olK) &=& \frac{1- \frac{\a^2}{4} (K+ \olK) -
                \sqrt{[1- \frac{\a^2}{4} (K+ \olK)]^2 - \frac{\a^4}{4}
                 K \olK }} {\frac{\a^4}{8}~ K \olK} = \non \\
&& \non \\
 &=& 1 + \frac{\a^2}{4} (K+ \olK) + {\cal O}(\a^4)
\eea

Since we are interested in performing quantum calculations
we rewrite the action in
(\ref{sbi2}) in $N=1$ superspace: there one can use the standard
quantization
procedure for
$N=1$ superfields. To this end we decompose the $N=2$
superfield-strength
${\calW}$ in terms of its $N=1$ superfield components
\bea
\calW &=& \F + \th_{\uld}^{~\a} W_\a - \th_{\uld}^{~2}  \olD^2 \olPhi
        + \frac i2 \th^\a_{\uld} \olth^\ad_{\uld} \partial_{\a\ad} \F +
                \non \\
        && + \frac i2 \th^2_{\uld} \olth^\ad_{\uld} \partial_{\a\ad} W^\a
        + \frac 14 \th^2_{\uld} \olth^2_{\uld} \Box \F
\eea
where $W_\a=i\olD^2 D_\a V$ is the $N=1$ chiral field-strength with $V$
the
unconstrained vector superfield, and $\Phi$ is the $N=1$ matter
superfield.
In this manner the kinetic part of the action in (\ref{sbi2})
becomes (we define $\th_{\ulu\a} = \th_\a$,
$\olth_{\ulu\ad} = \olth_\ad$ )
\bea
S_{KIN} &=&  \frac 12 \int d^4 x d^4 \th ~ \calW^2 =
                \non \\
        &=& \int d^4 x d^2 \th ~ W^2 +
        \int d^4 x d^2 \th d^2 \olth ~ \F \olPhi
\label{Skin}
\eea
The next contribution is given by the quartic interaction term
$\calW^2 \olcalW^2 $
which can
be easily decomposed in terms of $N=1$ superfields
\bea
\calW^2 \olcalW^2 \vert_{\th^2_{\uld} \olth^2_{\uld}}
        &=& \frac 14 \F^2 \Box \olPhi^2
                + \frac 14  \olPhi^2\Box\F^2
                +\frac 14 i\pa^{\a\ad}(\F^2) ~i\pa_{\a\ad}(\olPhi^2)
                        \non \\
&& + 4 \F \olD^2 \olPhi ~ \olPhi D^2 \F
        + 2 \F \olD^2 \olPhi ~\olW^\bd \olW_\bd +
                \non \\
&& + 2 W^\b W_\b \olPhi D^2 \F
        +  W_\b W^\b \olW_\bd \olW^\bd +
                \non \\
&& - 2i \F \olPhi W_\b \partial^{\b\bd} \olW_\bd
        - 2i \F \partial^{\b\bd} \olPhi W_\b  \olW_\bd +
                \non \\
&& + 2i \F \olPhi  \partial^{\b\bd} W_\b \olW_\bd
        + 2i W_\b \partial^{\b\bd} \F \olPhi ~ \olW_\bd
\label{WWbar}
\eea
The last two lines combine and up to total
derivative terms the ${\cal O}(\a^2)$ contributions to the
action are
\bea  \label{SBIN1}
S_{SBI}^{({\a}^2)}
&=&  \frac{\a^2}{8} \int d^4 x d^4 \th d^4\olth ~\calW^2 \olcalW^2  =
                        \non \\
&=& \frac{\a^2}{2} \int d^4 x d^2 \th d^2 \olth ~ W^2 \olW^2
- \frac{\a^2}{2} \int d^4 x d^2 \th d^2 \olth~ \olPhi~ \olW^\ad
i\pa_{\a\ad} \left(W^\a \F\right)
                        \non \\
&& + \frac{\a^2}{2} \int d^4 x d^2 \th d^2 \olth
        ~\left( \olW^2 \F \olD^2 \olPhi + W^2 \olPhi D^2 \F \right) +
                        \non \\
&& + \frac{\a^2}{4} \int d^4 x d^2 \th d^2 \olth
                ~\F i\pa^{\a\ad}\F\olPhi i\pa_{\a\ad} \olPhi +
                        \non \\
&&+ \frac{\a^2}{2} \int d^4 x d^2 \th d^2 \olth ~
        \F \olD^2 \olPhi ~\olPhi D^2 \F
\eea

The other interactions from the expansion of ${\calY(K,\olK) }$ can be
worked out straightforwardly, but the terms in (\ref{SBIN1}) are
sufficient for the determination of the one-loop counterterm.
The gauge-fixing for the vector superfield action is standard
(see for example \cite{superspace}). We work in Feynman gauge and use
the
following propagators
\bea
&& <V(x,\theta)V(x',\theta')>= \frac{1}{\Box} \d^{(4)}(x-x')
        \d^{(4)}(\theta-\theta') \non \\
&&<\Phi(x,\theta)\bar{\Phi}(x',\theta')> = - \frac{1}{\Box}
        \d^{(4)}(x-x') \d^{(4)}(\theta-\theta')
\label{propagators}
\eea
The vertices are obtained from (\ref{SBIN1}) with extra $\olD^2$
($D^2$)
factors acting on each chiral $\Phi$ (antichiral $\olPhi$)  line
leaving a vertex.

We are interested in the computation of ${\cal O}(\a^4)$
{\em on-shell}   one-loop divergent
contributions to the effective action  and in particular
we look for terms with external vector field-strengths. Once we
have determined the structure of these counterterms, $N=2$
supersymmetry will
allow us to reconstruct the complete answer. Setting the external
background
on-shell means that
we can freely use the equations of motion
\beq
D^\a W_\a = 0 \qquad \qquad \olD^\ad \olW_\ad = 0
\label{onshell}
\eeq
 which  imply
\beq
D^2 W_\a = - D_\a D^\b W_\b = 0 \qquad \qquad \olD^2\olW_\ad=0
\label{moreonshell}
\eeq

From the interaction terms in (\ref{SBIN1})
we see that the relevant one-loop supergraphs are the ones
with four external field-strengths and quantum propagators
given   by vector superfields and by matter chiral superfields.
First we consider the diagrams with vector internal lines.
Various different diagrams can be drawn:
most of
them are immediately discarded either because they trivially vanish
after
completion of the $D$-algebra, or are zero by momentum integration,
or else they reduce to tadpole-type
integrals which are set to zero in dimensional regularization scheme.
The only non vanishing
contributions  correspond to the supergraphs shown in fig. \ref{fig1}
and
\ref{fig2}
which we call $G_1$ and $G_2$ respectively in the following.

\bfg[htb] \bc
\mbox{\epsfig{figure={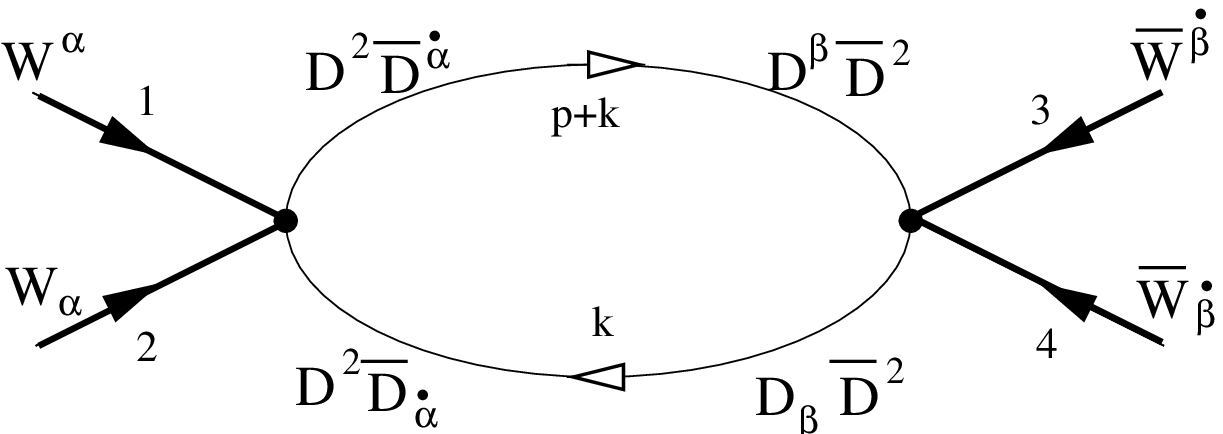},
       width=0.5 \linewidth}}
\caption{Supegraph $G_1$.}
\label{fig1}

\ec\efg

\bfg[htb] \bc
\mbox{\epsfig{figure={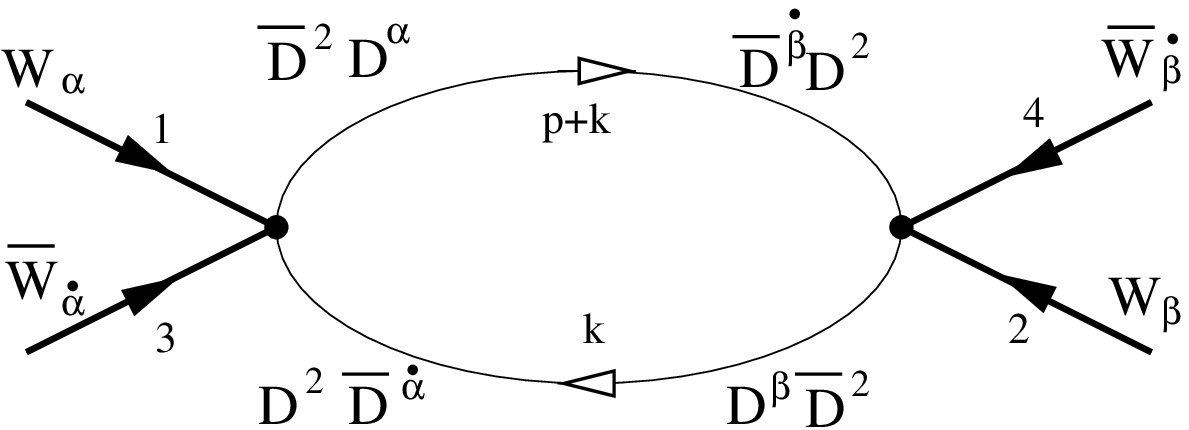},
      width=0.5 \linewidth}}
\caption{Supegraph $G_2$.}
\label{fig2}
\ec\efg

First we analyze the contribution to the effective action from $G_1$.
After $D$-algebra we obtain
\bea \label{primoG1}
\G^{(4)}_{G_1} \left[ W, \olW \right] &=& - \frac{\a^4}{32}
        \int d^4 p_1 \dots d^4 p_4 ~
        \d \left( \S p_i \right)  d^2 \th d^2\olth
        ~ D^2 \left[ W^\a (-p_1) W_\a (-p_2) \right]
        \\ \non
&&      \int \frac{d^{4-2\e} k}{(2\pi)^4}~
        \frac{(p+k)^{\b\bd} (-k)_{\b\bd}}{(p+k)^2 k^2}
        ~ \olD^2 \left[ \olW^\ad (-p_3) \olW_\ad (-p_4) \right]
\eea
where $p = p_1 + p_2 = - p_3 - p_4$ and momenta are labeled as shown in
fig. \ref{fig1}.

Using dimensional regularization the momentum integral in
(\ref{primoG1})
gives the following divergent contribution
\bea
&& \int \frac{d^{4-2\e} k}{(2\pi)^4} \frac{(p+k)^{\a\ad}
        ~ k_{\a\ad}}{(p+k)^2 k^2}
 \rightarrow - \frac{1}{\e} \frac{p^2 }{(4\pi)^2}
\label{G1div}
\eea
Therefore the divergence associated to the $G_1$ graph can be written
as
\bea
\G^{(4)}_{G_1~div.} \left[ W, \olW \right]
        &=& \frac{\a^4}{8} \left( - \frac{1}{\e}
        ~\frac{1}{(4\pi)^2} \right)\int d^4 p_1 \dots d^4 p_4
        \d \left( \S p_i \right)  d^2\th d^2\olth
                \non \\
&& (p_1+p_2)^2 W^2 (1,2)  D^2 \olD^2 \olW^2 (3,4)
\label{strano}
\eea
In order to streamline the notation we write
$W^2(1,2)\equiv \frac{1}{2} W^\a(-p_1) W_\a(-p_2)$ and also
$W_{\a,1}\equiv W_\a(-p_1)$ etc. Sometimes, whenever clear from the
context, we simply drop the momentum suffixes.

It is useful to rewrite the result in (\ref{strano})
taking advantage of various on-shell
identities. We list the ones that we repeatedly use
\bea
&& \olD^2 \olW^2 =  - \frac{1}{2} \left( \olD^\ad \olW^\bd \right)
                \left( \olD_\ad  \olW_\bd \right)\nonumber\\
&& D^2 \olD^\ad \olW^\bd = [D^2, \olD^\ad] \olW^\bd =
                - p^{\a\ad} D_\a \olW^\bd =0\nonumber\\
&&p^2 W_\a(p) = - \olD^2 D^2 W_\a = 0
\label{variousidentities}
\eea
With the help of the above relations we can write
\bea
&&D^2 \olD^2\olW^2(3,4)= - \frac{1}{2}
\left[ D^\a \olD^\ad \olW^\bd(-p_3) \right]
        \left[ D_\a \olD_\ad
        \olW_\bd(-p_4) \right]=\nonumber\\
&& = - p_3^{~\a\ad} p_{4~\a\ad}\olW^2(3,4) =
- (p_3 + p_4)^2 \olW^2(3,4)
\eea
The final result for the divergent contribution is
\bea
\G^{(4)}_{G_1~div.} \left[ W, \olW \right] &=&
        \frac{\a^4}{8}~ \frac 1\e~ \frac{1}{(4\pi)^2}~
        \int d^4 p_1 \dots d^4 p_4
        ~ \d \left( \S p_i \right)  d^2 \th d^2\olth
                \nonumber\\
&& \qquad s^2 W^2(1,2) \olW^2 (3,4)
\label{G1}
\eea
We have introduced the Mandelstam variable
$s = (p_1+p_2)^2 = (p_3+ p_4)^2$. In short we write
\bea
&& \G^{(4)}_{G_1~div.} \left( p_1 , \dots, p_4 \right) =
        \frac{\a^4}{8}~ \frac 1\e~ \frac{1}{(4\pi)^2}~ s^2
\label{div1G1}
\eea

\vspace{0.4cm}

\noindent
Now we turn to the corresponding calculation for the supergraph in fig.
\ref{fig2}.

\vspace{0.4cm}

After completion of the $D$-algebra
the $G_2$ diagram in fig. \ref{fig2} gives rise to
the following contribution to the
effective action
\bea
\G^{(4)}_{G_2} \left[ W, \olW \right] &=& \frac{\a^4}{8}
        \int d^4 p_1 \dots d^4 p_4 ~
        \d \left( \S p_i \right)  d^2 \th d^2\olth
        ~ \olD^\cd \left[ D^\c W_\a(-p_1) \olW_\ad(-p_3) \right]
                        \non \\
&& \qquad \int \frac{d^{4-2\e} k}{(2\pi)^4}
        \frac{(p+k)_{\c\cd}~(p+k)^{\a\bd}~ k^{\b\ad}}
        {(p+k)^2 k^2}
        W_\b (-p_2)~ \olW_\bd (-p_4)
\eea
where now we have set $p= p_1 + p_3 = - p_2 - p_4$.
The one-loop momentum integral leads to a divergence which in
dimensional regularization is given by
\newpage
\bea
&& \int \frac{d^{4-2\e} k}{(2\pi)^4}
        \frac{(p+k)_{\c\cd}~(p+k)^{\a\bd}~ k^{\b\ad}}
        {(p+k)^2 k^2}
                        \non  \\
&& \qquad \rightarrow \frac{1}{12\e} \frac{1}{(4\pi)^2}
        \left\{ -p_{\c\cd} p^{\a\bd} p^{\b\ad}
        - p_{\c\cd} ~ C^{\a\b}~ C^{\bd\ad} ~p^2  \right.
                \non  \\
&& \qquad \qquad \left.+ p^{\b\ad}   ~p^2
        ~ \d_\c^{~\a}~ \d_\cd^{~\bd}
                 - p^{\a\bd}  ~p^2
        ~ \d_\c^{~\b} ~ \d_\cd^{~\ad}  \right\}
\eea
so that we obtain
\bea
\G^{(4)}_{G_2~div} \left[ W, \olW \right] &=& \frac{\a^4}{96}
        ~\frac{1}{\e} ~\frac{1}{(4\pi)^2}
        \int d^4 p_1 \dots d^4 p_4
        ~\d \left( \S p_i \right)  d^2 \th d^2\olth
                        \non \\
&& \left\{-
        p_{\c\cd} p^{\a\bd} p^{\b\ad}
        [ \olD^\cd D^\c (W_{\a ,1} \olW_{\ad ,3}) ]
        W_{\b ,2} \olW_{\bd ,4} \right.
                        \non \\
&& + p_{\c\cd}  p^2
        [ \olD^\cd D^\c (W^\a_1 \olW^\ad_3) ]
        W_{\a ,2} \olW_{\ad ,4}
                        \non \\
&& + p^{\b\ad} p^2
        [ \olD^\bd D^\a (W_{\a ,1} \olW_{\ad ,3}) ]
        W_{\b ,2} \olW_{\bd ,4}
                        \non \\
&& \left. - p^{\a\bd}  p^2
        [ \olD^\ad D^\b (W_{\a ,1} \olW_{\ad ,3}) ]
        W_{\b ,2} \olW_{\bd ,4}  \right\}
\eea
Now we impose the on-shell equations
(\ref{onshell},\ref{moreonshell},\ref{variousidentities})
on the external background. In addition we use the following relations
\bea
&&  p_{\c\cd} (D^\c W_{\a ,1}) (\olD^\cd \olW_{\ad ,3}) =
        (p_1 + p_3)_{\c\cd}
        (D^\c W_{\a ,1}) (\olD^\cd \olW_{\ad ,3}) = 0
\eea
Also for two spinors $\psi_\a$, $\chi_\a$ we can write
\beq
\psi_\a \chi_\b = \frac{1}{2} \psi_{(\a} \chi_{\b)} + \frac{1}{2}
C_{\b\a}
        \psi^\c \chi_\c
\eeq
In this way we have
\bea
&& p^{\a\bd} p^{\b\ad} W_{\a ,1} ~\olW_{\ad ,3}
        ~ W_{\b ,2} ~\olW_{\bd ,4} =
                \non \\
&&=-p^{\a\bd} p^{\b\ad}\left[ \frac{1}{2}(W_{\a ,1} W_{\b ,2} +
        W_{\b ,1} W_{\a ,2})+\frac{1}{2} C_{\b\a}W^\c_1 W_{\c,2}\right]
                \nonumber\\
&& \qquad \qquad \left[ \frac{1}{2}(\olW_{\ad ,3} \olW_{\bd ,4}
        +\olW_{\bd ,3} \olW_{\ad ,4})+\frac{1}{2} C_{\bd\ad}
        \olW^\cd_3 \olW_{\cd ,4} \right]
                \non \\
&& = - \frac{1}{2} ~ p^{\a\bd} p^{\b\ad} W_{\a ,1} W_{\b ,2}
        \olW_{\ad ,3} \olW_{\bd ,4}
        - C_{\b\a} ~ C_{\bd\ad}~ W^2(1,2)~ \olW^2(3,4) ~p^{\a\bd} p^{\b\ad}
\label{newident}
\eea
since on-shell
\bea
&& p^{\a\bd} p^{\b\ad} W_{\a ,1} W_{\b ,2}
        \olW_{\bd ,3} \olW_{\ad ,4} =0
                \non \\
&&p^{\a\bd} p^{\b\ad}(W_{\a ,1} W_{\b ,2} +W_{\b ,1} W_{\a ,2})
        C_{\bd\ad}\olW^\cd_3 \olW_{\cd ,4}=0
\eea
Thus (\ref{newident}) leads to the on-shell identity
\bea
&& p^{\a\bd} p^{\b\ad} W_{\a ,1} ~\olW_{\ad ,3}
        ~ W_{\b ,2} ~\olW_{\bd ,4} =
                \non \\
&&=- 2~C_{\b\a} ~ C_{\bd\ad}~ W^2(1,2)~ \olW^2(3,4) ~p^{\a\bd}
p^{\b\ad}
                \non \\
&& = 4~ p^2 W^2(1,2)~\olW^2(3,4)
\eea
Finally we obtain
\bea
&& \G^{(4)}_{G_2~div} \left[ W, \olW \right] = - \frac{\a^4}{8}
        ~\frac{1}{\e} ~\frac{1}{(4\pi)^2} ~\frac{1}{12}
        \int d^4 p_1 \dots d^4 p_4
        ~\d \left( \S p_i \right)  d^2 \th d^2\olth
                        \nonumber\\
&& \qquad \left\{ p_{\c\cd} p^{\a\bd} p^{\b\ad} (-p_1^{\c\cd})
        - p_{\c\cd} C^{\a\b} ~ C^{\ad\bd} p^2 (-p_1^{\c\cd})
        + p^{\a\bd} p^2 (-p_1^{\b\ad})  \right\}
                        \nonumber\\
&& \qquad \qquad W_\a (-p_1) \olW_\ad (-p_3)
        W_\b (-p_2) \olW_\bd (-p_4) =
                        \non \\
&& = \frac{\a^4}{8}
        ~\frac{1}{\e} ~\frac{1}{(4\pi)^2} ~
        \int d^4 p_1 \dots d^4 p_4
        \d \left( \S p_i \right)  d^2 \th d^2\olth ~
    (p^2)^2~ W^2(1, 2) ~\olW^2(3,4)
\label{G2}
\eea
In short we write
\bea
&& \G^{(4)}_{G_2~div} \left( p_1 , \dots, p_4 \right) =
        \frac{\a^4}{8}~ \frac 1\e~ \frac{t^2}{(4\pi)^2}
\label{div2G1}
\eea
where $t = (p_1+p_3)^2 = (p_2 + p_4)^2$.

Now we study the contributions from the diagrams with chiral internal
lines. Setting the external vector background on-shell, the relevant
vertex from the action in (\ref{SBIN1}) is $\frac{\a^2}{2}\int d^4x
d^2\theta d^2\bar{\theta}~W_\a \olW_\ad \olPhi i\pa^{\a\ad}\F$.
Thus we obtain a third graph that we call $G_3$. It is
straightforward to realize that the D-algebra is the same as the one
for the $ G_2$ diagram shown in fig. 2, with  $\ad \leftrightarrow \bd$ in 
the loop.
Repeating exactly the same steps as for $G_2$, we obtain
\bea
&& \G^{(4)}_{G_3~div} \left[ W, \olW \right] = \frac{\a^4}{8}
        ~\frac{1}{\e} ~\frac{1}{(4\pi)^2} ~\frac{1}{12}
        \int d^4 p_1 \dots d^4 p_4
        ~\d \left( \S p_i \right)  d^2 \th d^2\olth
                        \nonumber\\
&& \qquad p_{\c\cd} C^{\b\a} ~ C^{\bd\ad} p^2 (-p_1^{\c\cd})
 W_\a (-p_1) \olW_\ad (-p_3)
        W_\b (-p_2) \olW_\bd (-p_4) =
                        \non \\
&& = \frac{\a^4}{24}
        ~\frac{1}{\e} ~\frac{1}{(4\pi)^2} ~
        \int d^4 p_1 \dots d^4 p_4
        \d \left( \S p_i \right)  d^2 \th d^2\olth ~
    (p^2)^2~ W^2(1, 2) ~\olW^2(3,4)
\label{G3}
\eea

The total one-loop divergence for
$W^2\olW^2$ is given by the sum of (\ref{G1}), (\ref{G2}) and (\ref{G3})
\bea
&& \G^{(4)}_{div}\left[ W, \olW \right]=
        \frac{\a^4}{8}~ \frac 1\e~ \frac{1}{(4\pi)^2}~
        \int d^4 p_1 \dots d^4 p_4
        \d \left( \S p_i \right)  d^2 \th d^2\olth \nonumber\\
&& \qquad \qquad (s^2 +\frac 43 t^2)  ~ W^2(1, 2) ~\olW^2(3,4)
\label{resultmoment}
\eea

The $N=2$ supersymmetrization of the result is immediate since we  have
shown that the term $W^2\olW^2$ appears in
 the full $N=2$ superspace integral of
$\calW^2 \olcalW^2$ (see eq. (\ref{WWbar})). In fact, using the
equations of motion for the vectors and for the matter fields
($D^2\Phi=0$, $\olD^2 \olPhi=0$), the only terms that
survive on-shell are
\beq
\frac{\a^2}{2}\int d^4x
d^2\theta d^2\bar{\theta}~\left( W^2 \olW^2+W_\a \olW_\ad \olPhi i\pa^{\a\ad}\F
-\frac 14 \olPhi^2 i\pa^{\a\ad}\F i\pa_{\a\ad}\F\right)
\label{actiononshell}
\eeq

The ${\cal O}(\a^4)$ counterterm given in (\ref{resultmoment}) can be rewritten in
configuration space
\bea
&& \G^{(4)}_{div}  =
        \frac{\a^4}{8}~ \frac 1\e~ \frac{1}{(4\pi)^2}
        \int  d^4 x d^2 \th d^2 \olth
                \nonumber       \\
&& \qquad \big( \partial^\m W^\a ~ \partial_\m W_\a ~
        \partial^\n \olW^\ad ~\partial_\n\olW_\ad +\frac 43
        ~\partial^\m W^\a ~ \partial^\n W_\a ~
        \partial_\m \olW^\ad ~\partial_\n\olW_\ad \big)
\label{resultconfig}
\eea

\vspace{0.6cm}

In order to make contact with the standard bosonic Born-Infeld action,
we express the result in component form. The $\theta$-expansion of the
superfield-strength is
\bea
W_\a (x, \th, \olth) &=&
        \l_\a + \th^\b f_{\a\b} - i \th^\b C_{\b\a} D' + i \th^2
        \partial_{\a\ad} \oll^\ad +
                \\      \non
&& + \frac{i}{2} \th^\b \olth^\bd \partial_{\b\bd} \l_\a +
        \frac{1}{4} \th^2 \olth^2 \Box \l_\a -
                 \frac{i}{2} \th^2 \olth^\bd \partial^\b_{~\bd}
        \left( f_{\a\b} -
        i C_{\b\a} D' \right)
\eea
\bea
\olW_\ad (x, \th, \olth) &=&
        \oll_\ad + \olth^\bd \left( \olf_{\ad\bd}
        + i C_{\bd\ad} D' \right)
        + i \olth^2 \partial_{\a\ad} \l^\a +
                \\      \non
&& + \frac{i}{2}  \olth^\bd \th^\b \partial_{\b\bd} \oll_\ad +
        \frac{1}{4} \th^2 \olth^2 \Box \oll_\ad -
                 \frac{i}{2} \olth^2 \th^\c \partial_\c^{~\bd}
        \left( \olf_{\ad\bd} + i C_{\bd\ad} D' \right)
\eea
where $f_{\a\b}$ (similarly $\olf_{\ad\bd}$) is related to $F_{\m\n}$
by
\bea
&& f_{\a\b} \equiv \frac 12~ (\s^{\m\n})_{\a\b}~ F_{\m\n}
                \non \\
&& (\s_{\m\n})_\a^{~\b} = - \frac 14 ~(\s_\m \tilde{\s}_\n -
                \s_\n \tilde{\s}_\m )_\a^{~\b}
                \non \\
&& \s_\m = (1, \vec{\s}) \qquad \qquad
        \tilde{\s}_\m = (1, - \vec{\s})
\eea
Thus we can easily obtain the
$\th^2 \olth^2$ coefficient of the component expansion of $W^2
\olW^2$
\bea
&& W^\a~W_\a ~\olW^\ad ~ \olW_\ad \vert_{\th^2~\olth^2} =
                \non \\
&& =   \frac{1}{4}~ \l^\a \l_\a \Box \left( \oll^\ad \oll_\ad \right)
        + \frac{1}{4} ~ \Box \left( \l^\a \l_\a \right) \oll^\ad \oll_\ad
        + \frac{1}{4} ~ i\partial^{\b\bd}\left( \l^\a \l_\a \right)    
        i\partial_{\b\bd}\left( \oll^\ad \oll_\ad \right)
                \non \\
&& \qquad + 2 i~ \partial^{\b\bd}\left[\l^\a \left( f_{\a\b} - i C_{\b\a}
        D' \right)\right] \oll^\ad \left( \olf_{\ad\bd} + i C_{\bd\ad} D'
        \right) +
                \non \\
&& \qquad + 2 i~ \partial^{\b\bd}\left[\oll^\ad \left( \olf_{\ad\bd} + 
      i C_{\bd\ad} D' \right)\right]
      \l^\a  \left( f_{\a\b} - i C_{\b\a}D'\right) -
                \non \\
&& \qquad -4 ~ \l^\a  \partial_{\a\ad} \oll^\ad ~
        \oll^\bd  \partial_{\b\bd} \l^\b +
                \non \\
&& \qquad + 2i ~ \l^\c  \partial_{\c\cd} \oll^\cd
        \left( \olf^{\ad\bd} \olf_{\ad\bd}  +2i ~\olf^\ad_{~\ad} D'
        -2 (D')^2 \right) +
                \non \\
&& \qquad + 2i \left( f^{\a\b} f_{\a\b}  -2i ~f^\a_{~\a} D'
        -2 (D')^2 \right) \oll^\cd  \partial_{\c\cd} \l^\c +
                \non \\
&& \qquad + \left( f^{\a\b} f_{\a\b}  -2i ~f^\a_{~\a} D'
        -2 (D')^2 \right)
        \left( \olf^{\ad\bd} \olf_{\ad\bd} +2i ~\olf^\ad_{~\ad} D'
        -2 (D')^2 \right)
\eea
This expression can be simplified using the equations of motion for
the auxiliary field $D'$. One can easily show that to all orders in
$\a$ an
on-shell solution is given by $D'=0$. In this way we obtain
\bea
&& W^\a~W_\a ~\olW^\ad ~ \olW_\ad \vert_{\th^2~\olth^2, D'=0} =
                \non \\
&& =   \frac{1}{4}~ \l^\a \l_\a \Box \left( \oll^\ad \oll_\ad \right)
        + \frac{1}{4} ~ \Box \left( \l^\a \l_\a \right) \oll^\ad \oll_\ad
        + \frac{1}{4} ~ i\partial^{\b\bd}\left( \l^\a \l_\a \right)    
        i\partial_{\b\bd}\left( \oll^\ad \oll_\ad \right)
                \non \\
&& \qquad + 2 i~ \partial^{\b\bd} \left( \l^\a f_{\a\b}\right)
        \oll^\ad \olf_{\ad\bd} +  2 i~ \partial^{\b\bd}\left(\oll^\ad  
        \olf_{\ad\bd} \right) \l^\a  f_{\a\b}-
                \non \\
&& \qquad -4 ~ \l^\a  \partial_{\a\ad} \oll^\ad ~
        \oll^\bd  \partial_{\b\bd} \l^\b +
                \non \\
&& \qquad + 2i ~ \l^\c  \partial_{\c\cd} \oll^\cd
        \olf^{\ad\bd} \olf_{\ad\bd} +
        2i f^{\a\b} f_{\a\b} \oll^\cd  \partial_{\c\cd} \l^\c +
                \non \\
&& \qquad + f^{\a\b} f_{\a\b} \olf^{\ad\bd} \olf_{\ad\bd}
\eea
Going back to (\ref{resultconfig}) we obtain
\bea
&& \G^{(4)}_{div} [ W, \olW ] =
        \G^{(4)}_{div} [ F^+ , F^- ] + \dots =
                \non \\
&& = \frac{\a^4}{8}~ \frac{1}{\e}~ \frac{1}{(4\pi)^2}
        \int  d^4 x \left[ \partial^\m (F^+)^{\rho\s} ~
        \partial_\m (F^+)_{\rho\s} ~
        \partial^\n (F^-)^{\rho'\s'} ~\partial_\n (F^-)_{\rho'\s'} +
        \right.
                \non    \\
&& \qquad \left. + \frac 43 ~\partial^\m (F^+)^{\rho\s} ~
        \partial^\n (F^+)_{\rho\s} ~
        \partial_\m (F^-)^{\rho'\s'} ~\partial_\n (F^-)_{\rho'\s'}
        \right] +\dots
\eea
where we have introduced the definitions
\bea
&& \tilde{F}_{\m\n} =
        \frac{1}{2} \e_{\m\n}^{~~\rho\s} ~ F_{\rho\s} \qquad \qquad
        F^{(\pm)} = \frac 12 ( F \pm i \tilde{F}) \label{F+-}
\eea
and used the relation
\bea
&& f^{\a\b} f_{\a\b} \olf^{\ad\bd} \olf_{\ad\bd} = (F^+)^2 (F^-)^2
\label{deff}
\eea
In (\ref{deff}) we have dropped total derivative terms that anyway
would not contribute upon $d^4x$ integration.

We can rewrite the $\G^{(4)}[F]$ contribution in momentum space: using
\beq
(F^+)^2 (F^-)^2= F^4 - \frac 14 (F^2)^2
\eeq
with
\bea
&&F {\tilde{F}}  =  F^{\m\n} \tilde{F}_{\m\n} \qquad \qquad
        F^2  = F^{\m\n} F_{\m\n}
                \non \\
&& F^4 = F^{\m\n} F_{\n\l}F^{\l\rho} F_{\rho\m}=
        \frac 14 (F \tilde{F})^2 + \frac 12 (F^2)^2
\label{moredef}
\eea
we obtain
\beq \label{gamom}
\G^{(4)}_{div}[F] = \frac{\a^4}{32\e}\frac 1{(4\p)^2}
\int d^4 p_1 \dots d^4 p_4~\d(\S p_i) (s^2 +\frac 43 t^2)
\left[ F^4 - \frac 14 (F^2)^2 \right]
\eeq
Finally, since the
factor $F^4 - \frac 14 (F^2)^2 $ is completely symmetric
in the $s,t,u$ variables, the result in (\ref{gamom}) can be rewritten
as
\beq \label{gamom1}
\G^{(4)}_{div}[F] =\frac 76~ \frac{\a^4}{48\e}\frac 1{(4\p)^2}
\int d^4 p_1 \dots d^4 p_4~\d(\S p_i) (s^2 + t^2+u^2)
\left[ F^4 - \frac 14 (F^2)^2 \right]
\eeq

It is interesting to note that terms like the one in
eq. (\ref{gamom1}) have been obtained in the past from the study of
scattering
amplitudes of vector fields in type IIB string theory on the $D3$-brane
\cite{HK} and in type I open string theory \cite{Sch,GW,AT}.
Indeed the result obtained from string theory is of the form (cfr.
\cite{HK})
\beq
\label{amp}
\frac{\G(s)\G(t)}{\G(1+s+t)}K(\e^{(i)},p^{(j)})
\eeq
where $\e^{(i)}$ are the four polarization vectors and $p^{(j)}$ are
the
four external momenta. $K$ is a kinematic factor corresponding to the
term
$\left[ F^4 - \frac 14 (F^2)^2 \right]$. As mentioned above this factor
is
completely symmetric in the external momenta. Expanding the
$\G$-functions
in eq. (\ref{amp}) and symmetrizing with respect to the momenta we
obtain
a term proportional to
\beq
(s^2 + t^2 + u^2)\left[ F^4 - \frac 14 (F^2)^2 \right],
\eeq
i.e. of the same form as in (\ref{gamom1}).

\vspace {0.6cm}

The result in (\ref{gamom1})
 gives the ${\cal O}(\a^4)$ counterterm of the bosonic Born-Infeld
action. On dimensional grounds it is easy to determine the general
structure
of the allowed counterterms at any order.
Taking into account the mass dimensions of the various factors
\beq
d^4x\sim m^{-4}\qquad \quad \a\sim m^{-2}\qquad \quad
\pa_\m\sim m\quad \quad F_{\m\n}\sim m^2
\label{dimension}
\eeq
and keeping in mind that divergences correspond to local expressions,
we have that the structure of the counterterms at $L$-loop order can be
written symbolically as
\beq
\frac{1}{\a^2} \int d^4 x~(\a^2 \pa^4)^L (\a F)^{2n}\qquad \qquad n>1
\label{counterterm}
\eeq
It is clear that in the presence of constant curvature $F$ no divergent
correction arises.  On the contrary, whenever $\pa F$ is not negligible
 divergences
might appear at any loops and to all orders in the fields. The number
of derivatives is fixed $\pa^{4L}$ at $L$ loops, while any number of
fields
can be produced since $\a F$ is of mass dimensions zero.

In the same way one can repeat this analysis for the
supersymmetric theory. It would be quite interesting to perform higher
order computations explicitly and determine the
counterterms exactly.

\vspace{1.2cm}
In a recent paper \cite{marina} similar issues have been addressed and
one-loop corrections have been computed
for the $D3$ brane action.

\medskip
\section*{Acknowledgments}

\noindent
We thank Renata Kallosh for having raised our interest on the
subject.\\
This work was
supported by the European Commission TMR program
ERBFMRX-CT96-0045, in which A. DG. and D. Z. are associated
to the University of Torino.

\end{document}